# *SU(3)* analysis of non-factorizable contributions to decays of Bottom mesons emitting two pseudoscalar mesons

Maninder Kaur* and R. C. Verma†
*Department of Physics, Punjabi University,*
*Patiala – 147002, India.*

Abstract

Two-body weak decays of bottom mesons into two pseudoscalar mesons are examined employing *SU(3)* flavor symmetry for the nonfactorizable matrix elements. Using certain measured Cabibo-favored modes, we fix the reduced matrix elements and predict the branching ratios of the Cabibbo-angle favored and suppressed decays of *B* and $B_s$ mesons.

*PACS No.*:13.25.Hw, 14.40. Nd, 13.30.Eg.

* maninderkaur@pbi.ac.in
† rcverma@gmail.com

## 1. INTRODUCTION

It is now fairly established that the naive factorization model does not explain the data on weak hadronic decays of charm and bottom mesons. On one hand large $N_c \rightarrow \infty$ limit, which apparently was thought to be supported by *D*- meson phenomenology [1-3], has failed to explain *B*-meson decays, as *B*-meson data clearly demands [4] a positive value of the $a_2$-parameter. Thus it is required to analysis such terms.

A lot of work has been done to study the nonfactorization contributions in the weak hadronic decays of *B*-decays during the past two decades. The nonfactorizable terms may appear for several reasons, like FSI rescattering effects and soft-gluons exchange around the basic weak vertex [5-8]. The rescattering effects on the outgoing mesons have been studied in detail for bottom meson decays. As the nonfactorizable contributions being non-perturbative cannot be calculated from the first principles, we employ the *SU(3)*-flavor symmetry, which is quite reliable for hadronic interactions to throw some light on these contributions. In an earlier work [9], using isospin *SU(2)* symmetry flavor, we have searched for a systematics in the nonfactorizable contributions for various decays of $B^-$ and $\bar{B}^0$-mesons involving isospin 1/2 and 3/2 final states. We observed that the nonfactorizable isospin 1/2 and 3/2 reduced amplitudes seem to bear a universal ratio for $\bar{B} \rightarrow \pi D / \rho D / \pi D^*$ decay modes. In order to realize the full impact of the flavor symmetry, and to relate decays of strange bottom meson with those of nonstrange bottom mesons and also to relate CKM-enhanced mode $(\Delta b = 1, \Delta C = 1, \Delta S = 0)$, with CKM-suppressed mode $(\Delta b = 1, \Delta C = 1, \Delta S = -1)$, we generalize our methodology to the *SU(3)* flavor symmetry for treatment of the nonfactorizable parts of corresponding weak decay amplitudes [1-3, 5].

## 2. WEAK HAMILTONIAN

We give QCD modified weak Hamiltonian for the CKM-enhanced mode $(\Delta b = 1, \Delta C = 1, \Delta S = 0)$,

$$H_w = \frac{G_F}{\sqrt{2}} V_{cb} V_{ud}^* \left[ c_1 \left(\bar{d}u\right)\left(\bar{c}b\right) + c_2 \left(\bar{c}u\right)\left(\bar{d}b\right) \right], \tag{1}$$

and for the CKM- suppressed mode $(\Delta b = 1, \Delta C = 1, \Delta S = -1)$,

$$H_w = \frac{G_F}{\sqrt{2}} V_{cb} V_{us}^* \left[ c_1 \left(\bar{s}u\right)\left(\bar{c}b\right) + c_2 \left(\bar{c}u\right)\left(\bar{s}b\right) \right], \tag{2}$$

where $\bar{q}_1 q_2 = \bar{q}_1 \gamma_\mu \left(1 - \gamma_5\right) q_2$ denotes color singlet $V\!-\!A$ Dirac current and the QCD coefficients [10-11] at bottom mass scale are

$$c_1 = 1.132, \qquad c_2 = -0.287. \tag{3}$$

Hadronic weak matrix elements of an operator, say $\left(\bar{d}u\right)\left(\bar{c}b\right)$, receives contributions from the operator itself and from its Fierz transformed form,

$$(\bar{d}u)(\bar{c}b) = \frac{1}{N_c}(\bar{c}u)(\bar{d}b) + \frac{1}{2}\left(\bar{c}\lambda^c u\right)\left(\bar{d}\lambda^c b\right), \tag{4}$$

where $\bar{q}_1 \lambda^c q_2 \equiv \bar{q}_1 \gamma_\mu \left(1 - \gamma_5\right) \lambda^c q_2$ represents the color octet current.

Finally the weak Hamiltonian (1) for CKM-enhanced mode acquires the following form:

$$H_w^{CF} = \frac{G_F}{\sqrt{2}} V_{cb} V_{ud}^* \left[ a_1 \left(\bar{c}b\right)_H \left(\bar{d}u\right)_H + c_2 H_w^8 \right], \tag{5}$$

for color flavored (CF) diagrams, and

$$H_w^{CS} = \frac{G_F}{\sqrt{2}} V_{cb} V_{ud}^* \left[ a_2 \left(\bar{d}b\right)_H \left(\bar{c}u\right)_H + c_1 \tilde{H}_w^8 \right], \tag{6}$$

for color suppressed (CS) diagrams, where

$$H_w^8 = \frac{1}{2}\sum_{a=1}^{8}\left(\bar{c}\lambda^a u\right)\left(\bar{d}\lambda^a b\right), \qquad \tilde{H}_w^8 = \frac{1}{2}\sum_{a=1}^{8}\left(\bar{d}\lambda^a u\right)\left(\bar{c}\lambda^a b\right). \tag{7}$$

The subscript $H$ represents the change from quark currents to hadron field operators [12]. Similar treatment can be done for the CKM- suppressed weak Hamiltonian (2).

The weak Hamiltonan thus generates factorizable (produced through color singlet weak current) and nonfactorizable parts (produced through color-octet weak currents) of the weak decay amplitudes, which can be expressed as

$$A^{tot}(\bar{B} \to P_1P_2) = A^f(\bar{B} \to P_1P_2) + A^{nf}(\bar{B} \to P_1P_2). \tag{8}$$

**2.1. Evaluation of Factorizable terms**

Following the standard procedure to calculate the factorizable part in CKM-enhanced and CKM-suppressed modes, we obtain spectator quark decay amplitudes as given in Table 1 and Table 2 respectively.

We calculate the factorization contributions numerically using $N_c = 3$ (real value), which fixes

$$a_1 = 1.02, \quad a_2 = +0.11. \tag{9}$$

Numerical inputs for the decay constants [12-13] are

$$f_D = (0.207 \pm 0.009)GeV, \quad f_\pi = (0.131 \pm 0.002)\,GeV,$$
$$f_K = (0.156 \pm 0.009)GeV. \tag{10}$$

Momentum dependence of the form- factors is taken as

$$F_0\left(q^2\right) = \frac{F_0(0)}{\left(1 - q^2/m_s^2\right)^n}, \tag{11}$$

where *n=1* for the monopole formula and pole mass is given by the scalar meson carrying the quantum number of the corresponding weak current [1-2], *i.e,*

$$m_s = 6.80 GeV \text{ for } (\bar{c}b) \text{ current,}$$

$$m_s = 5.89 GeV \text{ for } (\bar{s}b) \text{ current,}$$

$$m_s = 5.78 GeV \text{ for } (\bar{u}b) \text{ or } (\bar{d}b) \text{ current.}$$

Form- factors at $q^2 = 0$ have the following values based on [3, 13-15]

$$\begin{aligned} &F^{BK} = (0.34 \pm 0.00), \quad F^{BD} = (0.66 \pm 0.03), F^{B\pi} = (0.27 \pm 0.05), \\ &F^{B\eta} = (0.22 \pm 0.05), F^{B\eta'} = (0.23 \pm 0.05), \\ &F^{BsDs} = (0.67 \pm 0.01), F^{BsK} = (0.23 \pm 0.01), \\ &F^{Bs\eta} = (0.24 \pm 0.01), F^{Bs\eta'} = (0.36 \pm 0.02). \end{aligned} \qquad (12)$$

There are many other calculations for these form-factors and decay constants, such as light-cone sum rules [14, 16], perturbative QCD approach [15, 17-18], FAT factorization assisted topological- amplitude approach [10], lattice QCD [19-20], LFCQM [13]. Incidentally, their estimates reasonably match with (10) and (12).

We use the $\eta$- $\eta$' mixing angle $\theta_p = \theta_{ideal} - \phi_P$, and $\phi_P = -15.4^\circ$ [12]. Using these numerical inputs, we calculate factorizable contributions as given in col. 3 of Table 1 for *CKM*-enhanced decays and in col. 3 of Table 2 for CKM- suppressed decays and the corresponding branching fractions are given in col. 2 of Tables 3 and 4 respectively.

The choice of $N_c = 3$ seems to work for $B^- \to D^0\pi^-$ and $B^- \to D^0K^-$ which require constructive interference between *CF* and *CS* terms. However, other decays, specially those arising from the *CS* diagram, are experiencing large deviations from the experimental values. It is clear from the Table 3 and 4 that the factorization alone is not sufficient to explain the measured branching fractions, which thus require nonfactorizable terms.

### 2.2. Nonfactorizable terms

The nonfactorizable contributions arise through the Hamiltonian made up of color-octet currents, $\left\langle P_1 P_2 \left| H_w^8 \right| B \right\rangle$ and $\left\langle P_1 P_2 \left| \tilde{H}_w^8 \right| B \right\rangle$. Using the Wigner Eckart theorem, these are expressed in terms of the *SU(3)* Clebsch-Gordan *(C. G.)* coefficients and the reduced weak amplitudes

$$\left\langle P_1 P_2 \left| H_w^8 \right| B \right\rangle = \sum_{\gamma} (C.G.) \left\langle P_1\, P_2 \left\| H_w^8 \right\| B \right\rangle_{\gamma}, \tag{13}$$

and

$$\left\langle P_1 P_2 \left| \tilde{H}_w^8 \right| B \right\rangle = \sum_{\gamma} (C.G) \left\langle P_1\, P_2 \left\| \tilde{H}_w^8 \right\| B \right\rangle_{\gamma}, \tag{14}$$

where $\gamma$ represents different weak diagrams like *W*-external emission, *W*-internal emission, *W*-exchange, *W*-annihilation and *W*-loop.

However, we employ tensor approach to evaluate the *C.G.* Coefficients, where the matrix element $\left\langle P_1 P_2 \left| H_w^8 \right| B \right\rangle$ is considered as a weak spurion + $B \rightarrow P + P$ scattering process, whose general structure can be written in the *SU(3)* symmetry framework as

$$\begin{aligned}\left\langle P_1 P_2 \left| H_w^8 \right| B \right\rangle = & [a(B^m\, P_m^i\, P_n\,) + d(\,B^i\, P_n^m\, P_m\,)]\, H_i^n + \\ & [a'(B^m\, P_m^i\, P_n\,) + d'(B^i\, P_n^m\, P_m\,)]\, H_i^n,\end{aligned} \tag{15}$$

where $B^a$ denotes triplet of *B*-meson

$$B^a \equiv \left(B^-, \bar{B}^0, \bar{B}_s^0\right), \tag{16}$$

$P_m$ denotes triplet of charm mesons,

$$P_a = \left[D^0, D^+, D_s^+\right], \tag{17}$$

and $P_b^a$ denotes $3 \otimes 3$ matrix of bottomless and charmless octet mesons,

$$P_j^i = \begin{bmatrix} P_1^1 & \pi^+ & K^+ \\ \pi^- & P_2^2 & K^0 \\ K^- & \bar{K}^0 & P_3^3 \end{bmatrix}, \tag{18}$$

where

$$P_1^1 = \frac{\pi^0 + \eta \sin\theta_p + \eta' \cos\theta_p}{\sqrt{2}},$$
$$P_2^2 = \frac{-\pi^0 + \eta \sin\theta_p + \eta' \cos\theta_p}{\sqrt{2}},$$
$$P_3^3 = -\eta \cos\theta_p + \eta' \sin\theta_p.$$

$H_j^i$ represents transformation behavior of the weak Hamiltonian (1) and (2) through the following *SU(3)* decomposition:

$$1 \otimes 1 \otimes 3^* \otimes 3 = 8 \oplus 1. \tag{19}$$

Since all the quarks appearing in the Hamiltonian are different, singlet does not contribute. Choosing $H_2^1$ and $H_3^1$ components of the weak spurion in Hamiltonian (15), respectively for *CKM*- enhanced mode and *CKM*- suppressed modes, we obtain nonfactorizable contributions to various $\bar{B} \to PP$ decays, which are given in the col. 2 of Table 5 and 6, where QCD coefficient $c_2$ and $c_1$ have been appropriately multiplied for CF and CS terms coming from (6) and (7) respectively. It is to be pointed out that *W*-loop and *W*-annihilation diagrams do not contribute to the decays considered in this work.

There exists a straight correspondence between the terms appearing in (15) and various quark level processes. The terms, involving the coefficients *(-d + d')*, represent *W*-exchange diagrams. Other terms, having coefficient *(a + a'*) represent *W*-external emission, *(-a + a')* represent *W*-internal emission, collectively called, spectator quark decay like diagrams, where the bottomless quark in the parent *B*-meson flows into one of the final state mesons [7, 21]. For the sake of clarity, we make the following substitution

$$E = a + a', I = -a + a', \text{ and } X = -d + d', \tag{20}$$

corresponding to $W$-external emission, $W$-internal emission and $W$-exchange diagrams respectively. Consequently all the nonfactorizable terms involve only three reduced amplitudes to be fixed as shown in col.3 of Table 5 and 6.

As we know that the nonfactorizable contributions are not calculable exactly from theory at present, so these have to be estimated from the available data. In order to reduce the number of unknown parameters, we start with ignoring the $W$-exchange contribution, which has been found to be small. Specially, these may not play significant role, when $W$-emission processes are contributing dominantly to a particular decay.

In order to evaluate these parameters, we choose those decay modes which are free from the FSI phases and $W$-exchange phenomenon. Using the experimental value of branching fraction for $B(B^- \to D^0\pi^-) = (4.68 \pm 0.13)\times 10^{-3}$, and subtracting the corresponding factorizable part (given in Table 1) from its experimental decay amplitude (given in col. 5 of Table 7) with positive sign, we calculate

$$a' = -(0.05 \pm 0.02)\, GeV^3. \tag{21}$$

Similarly, using the experimental value for $B(\overline{B}_s^0 \to D_s^+\pi^-) = (3.00 \pm 0.23)\times 10^{-3}$, we get

$$E = a + a' = (1.05 \pm 0.02)\, GeV^3, \tag{22}$$

for positive sign of its amplitude, which leads to the following

$$I = -a + a' = (1.15 \pm 0.32)\, GeV^3, \tag{23}$$

Consequently, we calculate values of the nonfactorizable contributions for all the remaining decays, which are given in col.3. of Table 7 and 8 for CKM- enhanced and CKM- suppresses decays, respectively. We also rewrite the factorizable amplitude values in col. 2 of Table 7 and 8 for the sake of comparison with nonfactorizable contributions.

By adding both of these, we give our results for the total amplitudes in col. 4 of Table 7 and 8, which agree well with experimental amplitudes given in the last col. of these Tables.

It is to be noted that for the case of decays involving *CF* processes, nonfactorizable terms are small in comparison to the factorization terms. In contrast, for the decays appearing through the *CS* process, nonfactorizable terms are significantly large, bringing theory closer to the experiment.

Using the decay rate formula

$$\Gamma\left(\bar{B} \to P_1 P_2\right) = \left|\frac{G_F}{\sqrt{2}} V_{cb} V_{ud}^*\right|^2 \frac{p}{8\pi m_B^2} \left|A\left(\bar{B} \to P_1 P_2\right)\right|^2, \tag{24}$$

we calculate branching fraction for all the decays which are given in col. 2 of Table 9 and 10, which are in better agreement than the case of factorization alone (given in Tables 3 and 4).

### 3. RESULTS AND DISCUSSIONS

1) The branching fraction calculated by factorization alone overestimates for $B^- \to D^0 \pi^-$ and $\bar{B}^0 \to D^+ \pi^-$ decays. Though we have taken $B(B^- \to D^0 \pi^-)$ as input, our results for $B(\bar{B}^0 \to D^+ \pi^-)$ shows drastic improvement when nonfactorizable contributions are included.

2) For $\bar{B}^0 \to D^0 \pi^0$ decay, we notice that the factorization alone underestimates its branching fraction, where as our value has the right order of magnitude. Here, we remind that these decays have shown the presence of FSI,

$$
\begin{aligned}
A(\bar{B}^0 \to \pi^- D^+) &= \frac{1}{\sqrt{3}}\left[ A_{3/2}^{\pi D} e^{i\delta_{3/2}^{\pi D}} + \sqrt{2} A_{1/2}^{\pi D} e^{i\delta_{1/2}^{\pi D}} \right],\\
A(\bar{B}^0 \to \pi^0 D^0) &= \frac{1}{\sqrt{3}}\left[ \sqrt{2} A_{3/2}^{\pi D} e^{i\delta_{3/2}^{\pi D}} - A_{1/2}^{\pi D} e^{i\delta_{1/2}^{\pi D}} \right],\\
A(B^- \to \pi^- D^0) &= \sqrt{3} A_{3/2}^{\pi D} e^{i\delta_{3/2}^{\pi D}}.
\end{aligned} \tag{25}
$$

Following the FSI- phase independent analysis, sum of the branching fractions,

$$
\begin{aligned}
B\left(\bar{B}^0 \to D^+\pi^-\right) + B\left(\bar{B}^0 \to D^0\pi^0\right) &= \left(35.7 \pm 5.1\right)\times 10^{-4} \quad \textit{Theo.}\\
&= \left(29.4 \pm 1.4\right)\times 10^{-4} \quad \textit{Expt.}
\end{aligned} \tag{26}
$$

is in good agreement with the experiment.

3) We notice that factorization alone gives very small values of branching fraction for $\bar{B}^0 \to D^0\eta / D^0\eta'$ decays. However, our predictions, are in nice agreement within experimental errors, when nonfactorizable terms are included.

4) We predict branching fraction for $\bar{B}_s^0 \to D^0 K^0$ decay to be *9.5×10*$^{-4}$, which is 25 times larger than the prior value without nonfactorization. This presents a litmus test for our scheme.

5) In the case of *CKM*- suppressed decay mode, $B^- \to D^0 K^-$ remains in agreement with the experimental value, and our prediction for $\bar{B}^0 \to D^0 \bar{K}^0$ gets significantly improved in comparison to the case of factorization alone. In fact, we notice that this decay occur largely through the nonfactorizable terms.

6) It may be also be noted that similar to $\bar{B} \to D\pi$ decays, $\bar{B} \to D\bar{K}$ decays are also subjected to *FSI* as their final state have two different Isospin states $I = 0$ and $I = 1$, which can evolve differently at the *B*-meson mass scale. Using the isospin analysis, we express their decay amplitudes as,

$$A(\bar{B}^0 \to D^+K^-) = \frac{1}{\sqrt{2}}\left[A_1^{KD}e^{i\delta_1} + A_0^{KD}e^{i\delta_0}\right],$$
$$A(\bar{B}^0 \to D^0\bar{K}^0) = \frac{1}{\sqrt{2}}\left[A_1^{KD}e^{i\delta_1} - A_0^{KD}e^{i\delta_0}\right], \quad (27)$$
$$A(B^- \to D^0K^-) = \sqrt{2}A_1^{KD}e^{i\delta_1},$$

where $\delta_0$ and $\delta_2$ represent respective phases of $I = 0$ and 1 in the final state. Squaring and adding the first two equations, we calculate phase- independent sum of the branching fractions,

$$B\left(\bar{B}^0 \to D^+K^-\right) + B\left(\bar{B}^0 \to D^0\bar{K}^0\right) = \left(2.94 \pm 0.42\right)\times 10^{-4} \quad \textit{Theo.}$$
$$= \left(2.38 \pm 0.21\right)\times 10^{-4} \quad \textit{Expt.} \quad (28)$$

which shows better agreement within the errors.

7) Among the strange bottom mesons decays, our predicted branching fraction for $\bar{B}_s^0 \to D_s^+K^-$ is in very nice agreement with the experimental value. Here, we notice that the nonfactorizable contributions show destructive interference with the facto-rizable amplitude.

8) We have also predicted branching fractions for, $\bar{B}_s^0 \to D^0\eta / D^0\eta'$ decays, which are significantly larger than that of factorization alone. We also calculate branching fractions of, $\bar{B}^0 \to D^0\eta / D^0\eta'$ and $\bar{B}_s^0 \to D^0\eta / D^0\eta'$ decays for other values of the $\eta$-$\eta$' mixing angle, which are compared in Table 11. These can be tested in future experiment and present good tests of our analysis.

9) It is worth pointing out that the decays $\bar{B}^0 \to D_s^+K^-, \bar{B}_s^0 \to D^0\pi^0$ and $\bar{B}_s^0 \to D^+\pi^-$ remain forbidden, as we have not included the *W*- exchange process. Though *W*-exchange factorizable contributions are expected to be highly suppressed due to the helicity and color arguments, its nonfactorizable counterpart may become noticeable due to the soft gluon exchange. Fixing the parameter *X* with

$B\left(\bar{B}^0 \to D_s^+ K^-\right) = \left(0.27 \pm 0.05\right) \times 10^{-4}$, as input, we predict

$$B\left(\bar{B}_s^0 \to D^0 \pi^0\right) = \left(0.7 \pm 0.2\right) \times 10^{-6},$$
$$B\left(\bar{B}_s^0 \to D^+ \pi^-\right) = \left(1.4 \pm 0.3\right) \times 10^{-6}, \qquad (29)$$

which may also be checked in the future experiments.

**Table 1. Spectator decay amplitudes for the CKM-enhanced decays ($\Delta b = 1, \Delta C = 1, \Delta S = 0$) $\left(\times G_F/\sqrt{2}\, GeV^3\right)$**

| Decay Process | Factorizable – Amplitudes ($A^f$) | Factorizable - Amplitudes ($A^f$) |
|---|---|---|
| $B^- \to D^0 \pi^-$ | $a_1 f_\pi F_0^{BD}(m_\pi^2)(m_B^2 - m_D^2) + a_2 f_D F_0^{B\pi}(m_D^2)(m_B^2 - m_\pi^2)$ | 9.482 ± 0.421 |
| $\bar{B}^0 \to D^0 \pi^0$ | $-a_2 f_D F_0^{B\pi}(m_D^2)(m_B^2 - m_\pi^2)/\sqrt{2}$ | -0.558 ± 0.103 |
| $\bar{B}^0 \to D^+ \pi^-$ | $a_1 f_\pi F_0^{BD}(m_\pi^2)(m_B^2 - m_D^2)$ | 8.687 ± 0.395 |
| $\bar{B}^0 \to D^0 \eta$ | $a_2 f_D F_0^{B\eta}(m_D^2)(m_B^2 - m_\eta^2) \sin\theta_p/\sqrt{2}$ | 0.586 ± 0.016 |
| $\bar{B}^0 \to D^0 \eta'$ | $a_2 f_D F_0^{B\eta'}(m_D^2)(m_B^2 - m_{\eta'}^2) \cos\theta_p/\sqrt{2}$ | 0.322 ± 0.013 |
| $\bar{B}^0 \to D_s^+ K^-$ | 0 | 0 |
| $\bar{B}_s^0 \to D^0 K^0$ | $a_2 f_D F_0^{Bs\,K}(m_D^2)(m_{Bs}^2 - m_K^2)$ | 0.840 ± 0.030 |
| $\bar{B}_s^0 \to D_s^+ \pi^-$ | $a_1 f_\pi F_0^{Bs\,Ds}(m_\pi^2)(m_{Bs}^2 - m_{Ds}^2)$ | 8.759 ± 0.404 |

**Table 2. Spectator decay amplitudes for the CKM-suppressed decays ($\Delta$b = 1, $\Delta$C = 1, $\Delta$S = -1)** $\left(\times G_F/\sqrt{2}\,GeV^3\right)$

| Decay Process | Factorizable – Amplitudes ($A^f$) | Factorizable - Amplitudes ($A^f$) |
|---|---|---|
| $B^- \to D^0 K^-$ | $a_2 f_D F_0^{BK}(m_D^2)(m_B^2 - m_K^2) + a_1 f_K F^{BD}(m_K^2)(m_B^2 - m_D^2)$ | 2.718 ±0.116 |
| $\bar{B}^0 \to D^0 \bar{K}^0$ | $a_2 f_D F_0^{BK}(m_D^2)(m_B^2 - m_K^2)$ | 0.277 ± 0.040 |
| $\bar{B}^0 \to D^+ K^-$ | $a_1 f_K F_0^{BD}(m_K^2)(m_B^2 - m_D^2)$ | 2.439 ± 0.111 |
| $\bar{B}_s^0 \to D_s^+ K^-$ | $a_1 f_K F_0^{BsDs}(m_K^2)(m_{Bs}^2 - m_{Ds}^2)$ | 2.459 ± 0.114 |
| $\bar{B}_s^0 \to D^0 \pi^0$ | 0 | 0 |
| $\bar{B}_s^0 \to D^+ \pi^-$ | 0 | 0 |
| $\bar{B}_s^0 \to D^0 \eta$ | $-a_2 \cos\theta_p f_D F_0^{Bs\eta}(m_D^2)(m_{Bs}^2 - m_\eta^2)$ | -0.121 ± 0.009 |
| $\bar{B}_s^0 \to D^0 \eta'$ | $a_2 \sin\theta_p f_D F_0^{Bs\eta'}(m_D^2)(m_{Bs}^2 - m_{\eta'}^2)$ | 0.144 ± 0.010 |

**Table 3. Spectator quark decay branching fractions for the CKM-enhanced decays (Δb = 1, ΔC = 1, ΔS = 0)**

| Decay Process | Theoretical Branching- Fraction (Only factorization) ($\times 10^{-4}$) | Experimental Branching Fraction (Exp-Br) ($\times 10^{-4}$) [12] |
|---|---|---|
| $B^- \to D^0 \pi^-$ | 50.51 ± 4.49 | 46.8±1.3 |
| $\bar{B}^0 \to D^0 \pi^0$ | 0.17 ± 0.06 | 2.63±0.14 |
| $\bar{B}^0 \to D^+ \pi^-$ | 40.06 ± 3.69 | 25.2±1.3 |
| $\bar{B}^0 \to D^0 \eta$ | 0.18 ± 0.01 | 2.36±0.32 |
| $\bar{B}^0 \to D^0 \eta'$ | 0.052± 0.004 | 1.38±0.16 |
| $\bar{B}^0 \to D_s^+ K^-$ | 0 | 0.27±0.05 |
| $\bar{B}_s^0 \to D^0 K^0$ | 0.36 ± 0.03 | --- |
| $\bar{B}_s^0 \to D_s^+ \pi^-$ | 39.12 ± 3.60 | 30.0±2.3 |

**Table 4. Spectator quark decay branching fractions for CKM- suppressed decays ($\Delta b = 1, \Delta C = 1, \Delta S = -1$)**

| Decay Process | Theoretical Branching-Fraction in the absence of nonfactorization ($\times 10^{-4}$) | Experimental Branching Fraction (Exp-Br) ($\times 10^{-4}$) [12] |
|---|---|---|
| $B^- \to D^0 K^-$ | 4.10 ±0.35 | 3.63±0.12 |
| $\bar{B}^0 \to D^0 \bar{K}^0$ | 0.04 ± 0.01 | 0.52±0.07 |
| $\bar{B}^0 \to D^+ K^-$ | 3.12 ± 0.28 | 1.86±0.2 |
| $\bar{B}_s^0 \to D_s^+ K^-$ | 3.05 ± 0.28 | 2.27±0.19 |
| $\bar{B}_s^0 \to D^0 \pi^0$ | 0 | --- |
| $\bar{B}_s^0 \to D^+ \pi^-$ | 0 | --- |
| $\bar{B}_s^0 \to D^0 \eta$ | 0.0075 ± 0.0011 | --- |
| $\bar{B}_s^0 \to D^0 \eta'$ | 0.010 ±0.001 | --- |

**Table 5. Nonfactorizable contributions to $B \to PP$ decays**

**($\Delta$b = 1, $\Delta$C = 1, $\Delta$S = 0)** $\left(\times G_F/\sqrt{2}\, GeV^3\right)$

| Decay Process | NonFactorizable- Amplitude $A^{nf}$ | NonFactorizable- Amplitude $A^{nf}$ |
|---|---|---|
| $B^- \to D^0 \pi^-$ | $2a'(c_1 + c_2)$ | $(I+E)(c_1 + c_2)$ |
| $\bar{B}^0 \to D^0 \pi^0$ | $(a-a'-d+d')\, c_1/\sqrt{2}$ | $(-I+X)\,(1/\sqrt{2})\, c_1$ |
| $\bar{B}^0 \to D^+ \pi^-$ | $(a+a'-d+d')\, c_2$ | $(E+X)c_2$ |
| $\bar{B}^0 \to D^0 \eta$ | $(-a+a'-d+d')\sin\theta_p\; c_1/\sqrt{2}$ | $(I+X)\sin\theta_p\; c_1/\sqrt{2}$ |
| $\bar{B}^0 \to D^0 \eta'$ | $(-a+a'-d+d')\cos\theta_p\; c_1/\sqrt{2}$ | $(I+X)\cos\theta_p\; c_1/\sqrt{2}$ |
| $\bar{B}^0 \to D_s^+ K^-$ | $(-d+d')\, c_2$ | $X\, c_2$ |
| $\bar{B}_s^0 \to D^0 K^0$ | $(-a+a')\; c_1$ | $I\; c_1$ |
| $\bar{B}_s^0 \to D_s^+ \pi^-$ | $(a+a')\; c_2$ | $E\, c_2$ |

**Table 6. Nonfactorizable contributions to $B \to P\ P$ decays**

**($\Delta$b = 1, $\Delta$C = 1, $\Delta$S = -1)** $\left(\times G_F / \sqrt{2}\, GeV^3\right)$

| Decay Process | NonFactorizable-Amplitude $A^{nf}$ | NonFactorizable-Amplitude $A^{nf}$ |
|---|---|---|
| $B^- \to D^0 K^-$ | $2a'(c_1 + c_2)$ | $(I+E)\ (c_1 + c_2)$ |
| $\bar{B}^0 \to D^0\ \bar{K}^0$ | $(-a+a')\ c_1$ | $I\ c_1$ |
| $\bar{B}^0 \to D^+ K^-$ | $(a+a')\ c_2$ | $E\ c_2$ |
| $\bar{B}_s^0 \to D_s^+ K^-$ | $(a+a'-d+d')\ c_2$ | $(E+X)\ c_2$ |
| $\bar{B}_s^0 \to D^0 \pi^0$ | $(-d+ d')\ c_2/\sqrt{2}$ | $X\ c_2/\sqrt{2}$ |
| $\bar{B}_s^0 \to D^+ \pi^-$ | $(-d+ d')\ c_2$ | $X\ c_2$ |
| $\bar{B}_s^0 \to D^0 \eta$ | $((a- a')\cos\theta_p$ $+ (-d+d')\sin\theta_p/\sqrt{2})\ c_1$ | $(-I\cos\theta_p + X\sin\theta_p/\sqrt{2})c_1$ |
| $\bar{B}_s^0 \to D^0 \eta'$ | $(2(-a+ a')\sin\theta_p$ $+(-d+ d')\cos\theta_p/\sqrt{2})\ c_1$ | $(I\sin\theta_p + X\cos\theta_p/\sqrt{2})\ c_1$ |

**Table 7. Decay Amplitudes of CKM-Favored Decays**
**($\Delta$b = 1, $\Delta$C = 1, $\Delta$S = 0)** $\left(\times 10^{-2}\, G_F / \sqrt{2}\, GeV^3\right)$

| **Decay Process** | **Factorizable-Amplitude** $A^f$ | **NonFactorizable - Amplitude** $A^{nf}$ | **Total Theoretical Amplitude** $A^{Tot}$ | **Experimental –Amplitude** $\lvert A^{exp}\rvert$ |
|---|---|---|---|---|
| $B^- \to D^0\, \pi^-$ | 9.482 ± 0.421 | -0.355±0.127 | 9.126±0.438 | 9.126±0.127$^{\dagger}$ |
| $\bar{B}^0 \to D^0\, \pi^0$ | -0.558 ± 0.103 | 3.642±1.012 | 3.084±1.017 | 2.225±0.059 |
| $\bar{B}^0 \to D^+\, \pi^-$ | 8.687 ± 0.395 | -1.088±0.332 | 7.599±0.516 | 6.890±0.178 |
| $\bar{B}^0 \to D^0\, \eta$ | 0.586 ± 0.016 | -2.818±0.783 | -2.23±0.785 | 2.124±0.144 |
| $\bar{B}^0 \to D^0\, \eta'$ | 0.322 ± 0.013 | -2.307±0.640 | -1.98±0.641 | 1.652±0.096 |
| $\bar{B}^0 \to D_s^+\, K^-$ | 0 | 0 | 0 | 0.723±0.067 |
| $\bar{B}_s^0 \to D^0 K^0$ | 0.840 ± 0.030 | -5.151±1.432 | -4.31±1.432 | --- |
| $\bar{B}_s^0 \to D_s^+ \pi^-$ | 8.759 ± 0.404 | -1.088±0.332 | 7.67±0.523 | 7.670±0.294$^{\dagger}$ |

$^{\dagger}$inputs

**Table 8. Decay amplitudes of CKM-suppressed decays ($\Delta$b = 1, $\Delta$C = 1, $\Delta$S = -1)** $\left(\times 10^{-2}\, G_F \big/ \sqrt{2}\, GeV^3\right)$

| Decay Process | Factorizable-Amplitude $A^f$ | NonFactorizable - Amplitude $A^{nf}$ | Total Theoretical Amplitude $A^{Tot}$ | Experimental –Amplitude $\vert A^{exp}\vert$ |
|---|---|---|---|---|
| $B^- \to D^0 K^-$ | 2.718 ±0.116 | -0.081±0.029 | 2.637±0.119 | 2.557± 0.042 |
| $\bar{B}^0 \to D^0 \bar{K}^0$ | 0.277 ± 0.040 | -1.176±0.327 | -0.899±0.329 | 0.995± 0.067 |
| $\bar{B}^0 \to D^+ K^-$ | 2.439± 0.111 | -0.248±0.076 | 2.191± 0.149 | 1.883±0.101 |
| $\bar{B}_s^0 \to D_s^+ K^-$ | 2.459 ± 1.114 | -0.248±0.076 | 2.211± 0.134 | 2.122± 0.089 |
| $\bar{B}_s^0 \to D^0 \pi^0$ | 0 | --- | --- | --- |
| $\bar{B}_s^0 \to D^+ \pi^-$ | 0 | --- | --- | --- |
| $\bar{B}_s^0 \to D^0 \eta$ | -0.121 ± 0.009 | 0.745±0.207 | 0.624±0.208 | --- |
| $\bar{B}_s^0 \to D^0 \eta'$ | 0.144 ± 0.010 | -0.910±0.253 | -0.765±0.254 | --- |

**Table 9. Branching fraction of CKM-favored decays**
**($\Delta b = 1, \Delta C = 1, \Delta S = 0$)**

| **Decay Process** | ***Theoretical-Branching*** **($\times 10^{-4}$)** | ***Experimental - Branching*** **($\times 10^{-4}$) [12]** |
|---|---|---|
| $B^- \to D^0 \pi^-$ | 46.8 ± 1.3 | 46.8 ± 1.3† |
| $\bar{B}^0 \to D^0 \pi^0$ | 5.05 ± 3.33 | 2.63 ± 0.14 |
| $\bar{B}^0 \to D^+ \pi^-$ | 30.7 ± 4.2 | 26.8 ± 1.3 |
| $\bar{B}^0 \to D^0 \eta$ | 2.61 ± 1.83 | 2.36 ± 0.32 |
| $\bar{B}^0 \to D^0 \eta'$ | 1.99 ± 1.28 | 1.38 ± 0.16 |
| $\bar{B}^0 \to D_s^+ K^-$ | 0 | 0.22 ± 0.05 |
| $\bar{B}_s^0 \to D^0 K^0$ | 9.5 ± 5.2 | --- |
| $\bar{B}_s^0 \to D_s^+ \pi^-$ | 30.0 ± 2.3 | 30.0 ± 2.3† |

†inputs

**Table 10. Branching fraction of CKM-suppressed decays ($\Delta b = 1, \Delta C = 1, \Delta S = -1$)**

| Decay Process | *Theoretical-Branching* ($\times 10^{-4}$) | *Experimental-Branching* ($\times 10^{-4}$) [12] |
|---|---|---|
| $B^- \to D^0 K^-$ | 3.86 ± 0.35 | 3.63 ± 0.12 |
| $\bar{B}^0 \to D^0 \bar{K}^0$ | 0.42 ± 0.31 | 0.52 ± 0.07 |
| $\bar{B}^0 \to D^+ K^-$ | 2.52 ± 0.31 | 1.86 ± 0.20 |
| $\bar{B}_s^0 \to D_s^+ K^-$ | 2.46 ± 0.30 | 2.27 ± 0.19 |
| $\bar{B}_s^0 \to D^0 \pi^0$ | --- | --- |
| $\bar{B}_s^0 \to D^+ \pi^-$ | --- | --- |
| $\bar{B}_s^0 \to D^0 \eta$ | 0.20 ± 0.13 | --- |
| $\bar{B}_s^0 \to D^0 \eta'$ | 0.29 ± 0.18 | --- |

**Table 11. Branching Fractions ( $\times 10^{-4}$) of $\eta / \eta'$ emitting Decays including Nonfactorization Terms**

| Decay Process | $\phi = -10.4^\circ$ | $\phi = -15.4^\circ$ | $\phi = -24.4^\circ$ | Expt-Branching [12] |
|---|---|---|---|---|
| $\bar{B}^0 \to D^0 \eta$ | 2.23 ± 1.56 | 2.61 ± 1.83 | 3.24 ± 2.28 | 2.36 ± 0.32 |
| $\bar{B}^0 \to D^0 \eta'$ | 2.43 ± 1.57 | 1.99 ± 1.28 | 1.27 ± 0.82 | 1.38 ± 0.16 |
| $\bar{B}_s^0 \to D^0 \eta$ | 0.24 ± 0.16 | 0.20 ± 0.13 | 0.13 ± 0.08 | --- |
| $\bar{B}_s^0 \to D^0 \eta'$ | 0.25 ± 0.16 | 0.29 ± 0.18 | 0.36 ± 0.24 | --- |